\documentclass[twocolumn,showpacs,pra,amsmath,amssymb,floatfix]{revtex4-1}
\input{epsf}

\usepackage{graphicx}
\usepackage{dcolumn}
\usepackage{bm}

\begin{document}
\title{Relaxation dynamics in an isolated long-range Ising chain}
\author{H. T. Ng and  Jing-Ning Zhang}
\affiliation{Center for Quantum Information, Institute for Interdisciplinary Information Sciences, Tsinghua University, Beijing 100084, P. R. China}

\date{\today}

\begin{abstract}
We consider a chain of trapped ions to interact with each other via long-range interactions. This system can be used to simulate the long-range Ising model.
We study the dynamics of quantum coherence of a single spin in the chain, where the spins are initially prepared in their upper states. The relaxation dynamics exhibits due to the genuine long-range interaction. The degree of quantum coherence of a single spin rapidly decreases and vanishes in the steady state.  However, our numerical result suggests that the conventional spin chain model, which truncates the interactions between the distant spins, cannot show the relaxation dynamics. This implies that the usual truncation in approximating the long-range interaction is not applicable to describing 
the non-equilibrium dynamics. The effect of the interaction range on the relaxation dynamics is studied. The higher relaxation rate will show if a system has a longer range of interaction. However, it takes a longer relaxation time in the vicinity of infinite interaction range. We also examine the dynamics of quantum coherence of a block of spins. Our result may shed light on the relationship between long-range interaction and the coherence dynamics of a quantum many-body system.
\end{abstract}


\maketitle

\section{Introduction}
Recently, the long-range Ising model have been realized
by using a string of trapped ions \cite{Friedenauer,Kim,Islam} and a two-dimensional crystal lattice \cite{Britton}, respectively. The interactions between ions can be engineered such that the range of interaction becomes tunable. This opens the possibilities to observe new phenomena \cite{Laflorencie,Sandvik} of a many-body system with long-range interaction. More recently, the violation of Lieb-Robinson bound \cite{Hauke} has been showed in a chain of trapped ions \cite{Richerme,Jurcevic} with variable-range interactions, where the Lieb-Robinson bound \cite{Lieb} governs the speed of information propagation in a one-dimensional system with finite-range interaction.  

Non-equilibrium dynamics of a closed many-body system is important in quantum statistical mechanics \cite{Polkovnikov,Eisert,Langen}. There are 
open questions remained unsolved \cite{Gogolin,Deutsch,Srednicki,Rigol,Goold} in this area. For instance, there is still no general bound on how fast the equilibration takes place \cite{Goold}. In fact, a subsystem of a many-body system will lose its quantum coherence \cite{Banuls,Ng} even if the entire system is in a pure state. It is natural to ask how the long-range interaction affects the relaxation process of a local subsystem. It may be useful to understand the role of
interaction in the relaxation dynamics. Additionally, quantum coherence \cite{Baumgratz} is of the essence to the applications of quantum information science \cite{Nielsen} and photosynthesis for efficient energy conversion \cite{Romero}. The study of the relationship between quantum coherence and long-range interaction provides insight into the behavior of coherence in a strongly interacting many-body system and its potential applications.

Indeed, the long-range Ising model has been studied extensively \cite{Koffel,Feig,Schachenmayer,Gong}.
In this paper, we study the dynamics of quantum coherence
of a single spin and a block of spins in a chain of trapped ions, where the ions interact with each other  through long-range interactions. Since the interaction range between ions can be tuned by appropriately adjusting the laser beams, the relationship of coherence dynamics and interaction range can be studied experimentally. Initially, all ions are prepared in their upper spin states. The degree of quantum coherence of a single spin rapidly decreases and then drops to nearly zero in the steady state. The relaxation dynamics exhibits in this long-range Ising model.

Conventionally, the long-range interaction
is approximated by keeping the interaction between a few neighboring spins only \cite{Sachdev}. Since the interaction between the distant neighbors is weak, this approximation provides a reasonable good estimation of the ground-state energy of the exact long-range model. It should be noticed that this approximation must be reexamined if the dynamics involves a lot of eigenstates. We compare the dynamics of the exact model with the ``approximate'' model. Our numerical examples show that the ``approximate'' model cannot exhibit the relaxation dynamics. But the short-time dynamics of the ``approximate'' model resembles the
exact model if the interactions between more distant neighbors are included. 

In addition, we examine the effect of interaction range
on the relaxation dynamics. If the range of interaction
increases, then the relaxation rate becomes higher. However, when the interaction range is close to infinity, the dynamics of quantum coherence behaves very differently. It takes a longer time to lose the coherence. The system, with the infinite range of interaction, does not relax.  We also numerically study the quantum coherence of a block of spins in a chain. The larger size of a block of spins indeed gives a higher degree of quantum coherence. The different sizes of block spins have the similar rates of relaxation, and the coherence becomes steady in a long time.

This paper is organized as follows: In Sec.~II, we introduce the long-range Ising model. In Sec.~III, we first introduce the definition on how to quantify quantum coherence. We discuss the numerical results of the dynamics of quantum coherence of a single spin in the chain. We then compare the dynamics of the exact long-range Ising model with the ``approximate'' models. We also discuss the relaxation dynamics of a block of spins in a chain. We provide a discussion in Sec.~IV and close our paper with a conclusion in Sec~V.

\section{System}
We consider a chain of trapped ions in a linear trap.
The two internal states of an ion form a spin. For example, either two hyperfine states in ${}^{171}{\rm Yb}^+$ \cite{Richerme} or two electron levels in ${}^{40}{\rm Ca}^+$ \cite{Jurcevic} can be used. The interactions between the spins can be produced by off-resonantly coupling the two internal states to the collective motion of the chain in the perpendicular direction via a laser beam \cite{Richerme,Jurcevic}. This forms a 1D spin chain with long-range interaction. The Hamiltonian of a chain of trapped ions, with long-ranged interaction, is written as ($\hbar=1$) \cite{Porras}
\begin{eqnarray}
H=\sum_{i<j}J_{ij}\sigma^x_{i}\sigma^x_{j},
\end{eqnarray}
where $\sigma^x_j$ are Pauli spin operators,
$J_{ij}=J/|i-j|^\alpha$ and $\alpha$ ranges between 0 and 3. We denote the states $|1_x\rangle_i$ and $|0_x\rangle_i$ as the eigenstates of spin-up and -down states
of ion $i$ in the $x$-direction.

Let us consider the $k$-th eigenstate $|E_k\rangle$ of the Hamiltonian $H$ which can be written as the product
states of all ions, i.e., $\prod_i|\sigma_i\rangle_i$ and $|\sigma_i\rangle_i$ is the eigenstate of ion $i$ in the $x$-direction and $\sigma_i$ is either equal to zero or one. The eigenenergy $E_k$ of the $k$-th eigenstate $|E_k\rangle$ is given by
\begin{eqnarray}
E_k&=&\sum_{i<j}J_{ij}(2\sigma_i-1)(2\sigma_j-1),
\end{eqnarray}
where $J_{ij}$ is positive.

To prepare the initial state, we consider a large magnitude of the transverse magnetic field to be applied to the ions. The Hamiltonian of the system is written as
\begin{eqnarray}
\label{Ham_B}
H'=\sum_{i<j}J_{ij}\sigma^x_{i}\sigma^x_{j}-B\sum_i\sigma^z_i,
\end{eqnarray}
where $B$ is the strength of the transverse field. 
Initially, the magnitude $|B|$ is much larger than $J_{ij}$. The system is prepared in a product state of ions in the spin-up states as
\begin{eqnarray}
\label{zinitial}
|\Psi(0){\rangle}&=&\prod^N_{i=1}|1_z\rangle_i,
\end{eqnarray}
where ion $i$ is in the eigenstate $|1_z\rangle_i$ in the
$z$-direction. Indeed, it is the ground state of the Hamiltonian in Eq.~(\ref{Ham_B}), where $|B|\gg{J_{ij}}$. The initial state can be expressed in terms of the eigenstates $|E_k\rangle$ as
\begin{figure}[ht]
\centering
\includegraphics[height=2.0cm]{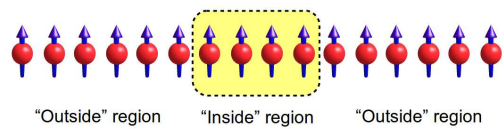}
\caption{ \label{LRising} (Color online) 
A chain of ions with long-range interaction. The system
is divided into two regions which are ``inside'' and ``outside'' subsystems, respectively.
}
\end{figure}
\begin{eqnarray}
|\Psi(0)\rangle&=&\prod^N_{j=1}\frac{1}{\sqrt{2}}(|0_x\rangle_j+|1_x\rangle_j),\\
\label{initial_state}
&=&\frac{1}{2^{N/2}}\sum^{2^{N}}_{k=1}|E_k\rangle.
\end{eqnarray}
It is a superposition of 
all eigenstates $|E_k\rangle$ with an equal weight.
We consider that the magnetic field is suddenly ramped down to zero. This initiates the system to become out of equilibrium.

\section{Quantum coherence Dynamics}
We consider the system to be divided into two subsystems which are ``inside'' and ``outside'' regions, respectively, as shown in Fig.~\ref{LRising}. The ``inside'' system can be described by the reduced density matrix $\rho^{\rm in}={\rm Tr}_{\rm out}(\rho)$, where $\rho^{\rm in}$ is obtained by tracing out the part of ``outside'' system. 

We study the dynamics of quantum coherence of spins in 
the ``inside'' system. To quantify quantum coherence, the coherence factor $C$ can be used. It is defined as \cite{Baumgratz}
\begin{eqnarray}
C&=&\sum_{i{\neq}j}|\rho^{\rm in}_{i,j}|, 
\end{eqnarray}
where $\rho^{\rm in}$ is the reduced density matrix of the system. The coherence factor $C$ is the sum of absolute value of all off-diagonal elements in the reduced density matrix $\rho^{\rm in}$. This quantity provides a proper measure of coherence of a system \cite{Baumgratz}.

\subsection{Relaxation of a single spin}
We first investigate the dynamics of coherence of a single spin in the chain and the other ions are treated as the ``outside'' system. The system of ion $j$ can be described
by the reduced density matrix as
\begin{eqnarray}
\rho^{\rm in}_j&=&\frac{1}{2}\left(\begin{matrix}
1 & f(t) \\
f(t)^* & 1
\end{matrix}\right).
\end{eqnarray}
The coherence factor $C$ is given by
\begin{eqnarray}
C(t)=|f(t)|.
\end{eqnarray}

To calculate the reduced density matrix $\rho^{\rm in}_j$, the part of ``outside'' region of the density matrix $\rho=|\Psi(t)\rangle\langle\Psi(t)|$ is traced out. From Eq.~(\ref{initial_state}), the state vector $|\Psi(t)\rangle$ is given by 
\begin{eqnarray}
|\Psi(t)\rangle&=&\frac{1}{2^{N/2}}\sum^{2^N}_{k=1}e^{-iE_k{t}}|E_k\rangle,
\end{eqnarray}
where the eigenstate $|E_k\rangle$ can be
written as
\begin{eqnarray}
|E_k\rangle&=&\underbrace{\prod_{j_k}|\sigma_{j_k}\rangle_{j_k}}_\text{``inside''}\underbrace{\prod_{i_k\neq{j_k}}|\sigma_{i_k}\rangle_{i_k}}_\text{``outside''}.
\end{eqnarray}
For the case of a single spin, the number of the states $|\sigma_{j_k}\rangle_{j_k}$ in the ``inside'' region to be equal to two, i.e., $|0\rangle$ and $|1\rangle$.

The density matrix is written as 
\begin{eqnarray}
\rho(t)&=&|\Psi(t)\rangle\langle\Psi(t)|,\\
&=&\frac{1}{2^N}\sum^{2^N}_{k=1}\sum^{2^N}_{k'=1}e^{-i(E_k-E_{k'}){t}}|E_k\rangle\langle{E_{k'}}|,\\
&=&\frac{1}{2^N}\sum^{2^N}_{k=1}\sum^{2^N}_{k'=1}e^{-i(E_k-E_{k'}){t}}\underbrace{|\sigma_{j_k}\rangle_{j_k}{}_{j_{k'}}\langle\sigma_{j_{k'}}|}_\text{``inside''}\nonumber\\
&&\times\underbrace{\prod_{i_k,i_{k'}\neq{j_k,j_{k'}}}|\sigma_{i_k}\rangle_{i_k}{}_{i_{k'}}\langle\sigma_{i_{k'}}|}_\text{``outside''}
\end{eqnarray}
If the ``outside'' part is traced out, the elements of 
density matrix can be retained for the two eigenstates $E_k$ and $E_{k'}$ with the same states of ``outside'' part, i.e., $\prod_{i_{k}\neq{j_{k}}}|\sigma_{i_k}\rangle_{i_k}$. In contrast, the elements of density matrix become zero if the two eigenstates $E_k$ and $E_{k'}$ with the different states of ``outside'' part. 

\begin{figure}[ht]
\centering
\includegraphics[height=8.5cm]{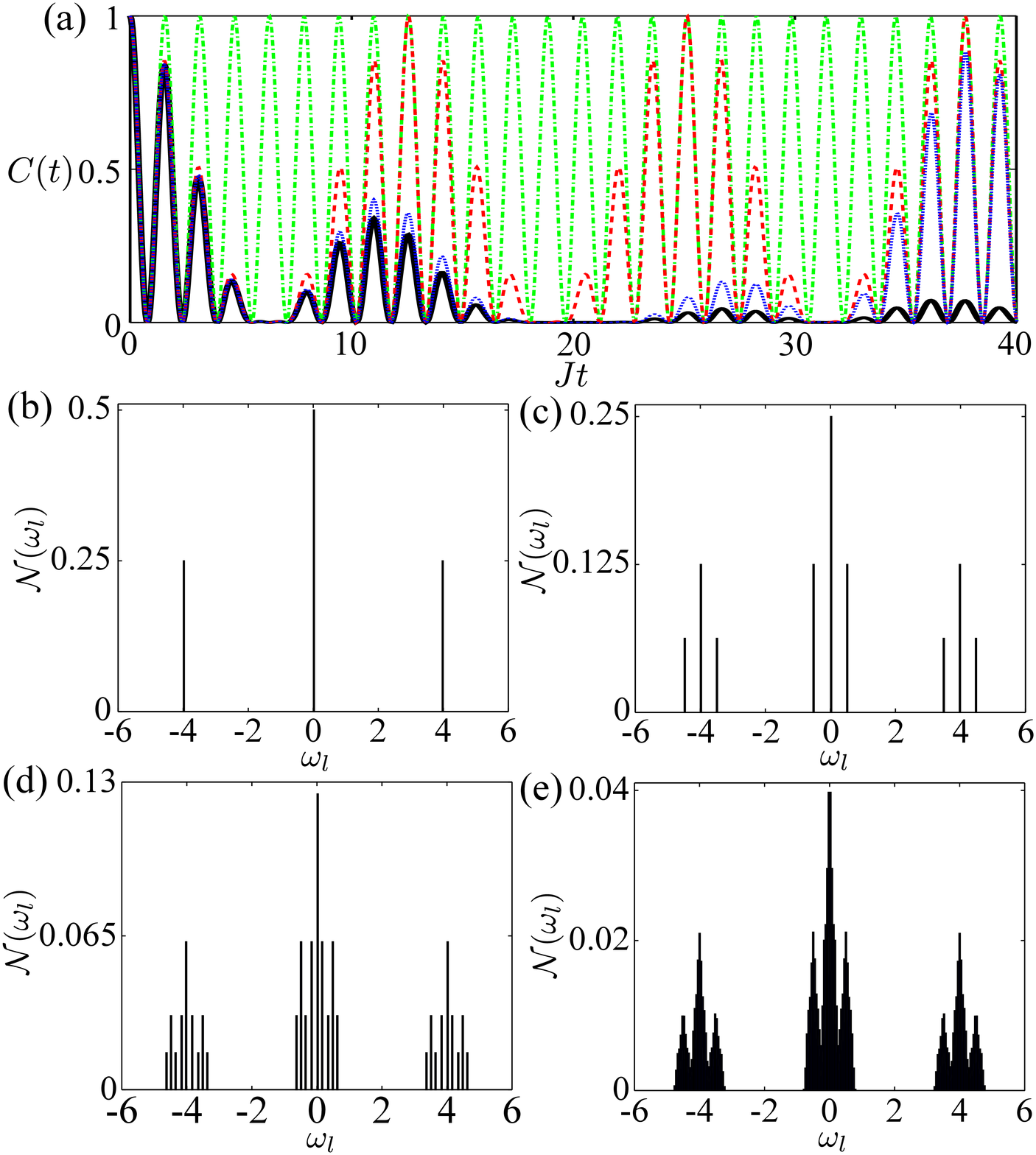}
\caption{ \label{csion} (Color online) In (a), coherence
factor $C(t)$ versus time $Jt$, for the ``approximate'' and exact long-range models, $N=20$ and $\alpha=3$. The coherence factor of the tenth ion is plotted.
The different types of interaction are denoted by the different lines: exact long-range interaction (black solid), next-next-nearest neighbor (blue dotted), next-nearest neigbor (red dashed), and nearest neighbor (green dot-dashed), respectively. Histograms of the effective frequencies $\omega_l$ are shown, where $\mathcal{N}(\omega_l)$ is the number frequency of $\omega_l$ and is normalized to be one. The different types of interaction in the different figures: (b) nearest neighbor, (c) next-nearest neighbor, (d) next-next-nearest neighbor and (e) exact long-range interaction without truncation, respectively.
}
\end{figure}

The reduced density matrix $\rho^{\rm in}_j$ can be obtained by summing up all possible $\tilde{\rho}^{\rm in}_l$, where the reduced density matrices $\tilde{\rho}^{\rm in}_l$ have the same
state of ``outside'' part, and $l=1,\ldots,2^{N-1}$. 
The density matrix $\tilde{\rho}^{\rm in}_l$ can be explicitly written as
\begin{eqnarray}
\tilde{\rho}^{\rm in}_l&=&\frac{1}{2^{N}}\left(\begin{matrix}
1 & e^{i\omega_l{t}} \\
e^{-i\omega_l{t}} & 1
\end{matrix}\right),
\end{eqnarray}
where $\omega_l=E_k-E_{k'}$ is the effective frequency,
and $E_k$ and $E_{k'}$ are the eigenenergies of the eigenstates $|E_k\rangle$ and $|E_{k'}\rangle$ with the same states of ``outside'' region. Hence, the reduced density matrix $\rho^{\rm in}_j$ is given by
\begin{eqnarray}
\rho^{\rm in}_j&=&\sum^{2^{N-1}}_{l=1}\tilde{\rho}^{\rm in}_l,\\
&=&\frac{1}{2}\left(\begin{matrix}
1 & f(t) \\
f^*(t) & 1
\end{matrix}\right),
\end{eqnarray}
where 
\begin{eqnarray}
f(t)=\frac{1}{2^{N-1}}\sum^{2^{N-1}}_{l=1}e^{i\omega_l{t}}.
\end{eqnarray}
The coherence factor $C$ is given by
\begin{eqnarray}
\label{coherence_factor}
C(t)&=&\frac{1}{2^{N-1}}\Bigg|\sum^{2^{N-1}}_{l=1}e^{i\omega_l{t}}\Bigg|.
\end{eqnarray}

It is interesting to compare the relaxation dynamics of  the exact long-range Ising model with
the ``approximate'' model, where the interactions between distant neighbors are truncated in the ``approximate'' one. Indeed, it is an usual approximation by including the interaction with a few neighbors only in the conventional spin chain models and the long-range interaction between the distant neighbors is neglected. 

Here we study the dynamics of quantum coherence of a single spin at the centre of a chain. In Fig.~\ref{csion}(a), we plot the coherence factor $C(t)$ versus time, for the "approximate" and exact long-range models. For the exact long-range model, the coherence factor $C$ rapidly decreases and then several collapses and revivals occur with decreasing magnitude. On the contrary, the coherence factors $C$ oscillate periodically if the ions interact with their nearest neighbors only. The collapses and rivials are shown if the ions interact with their next-nearest and next-next-nearest neighbors, but their magnitudes do not decrease in a longer time.

It should be noticed that if the interactions between more distant neighbors are included, then the short-time behaviors resemble the exact case in Fig.~\ref{csion}(a). The revivals of coherence appear subsequently, but they differ in a longer time. These numerical examples suggest that the "approximate" model with truncation of long-range interaction between distant neighbors cannot display the relaxation dynamics.

From Eq.~(\ref{coherence_factor}), the coherence factor is a function of a sum of effective frequencies $\omega_l$. In fact, the distribution of the effective frequencies $\omega_l$ can indicate the important feature of the coherence dynamics. In Figs.~\ref{csion}(b)-(e), we plot the histograms of the effective frequencies $\omega_l$, for the exact model and the different types of ``approximation'' of long-range interaction. For the case of the nearest-neighbor interaction, the effective frequencies are equal to three different values only in Fig.~\ref{csion}(b). As the interactions with more neighboring spins are included, a few effective frequencies distribute around the three different values. For the exact model, the three main peaks are shown and the effective frequencies smoothly spread around the peaks. This means that a lot of different effective frequencies $\omega_l$ contribute in the dynamics by using the exact model. The net effect of the different frequencies cancel out with each other, and therefore it leads to the loss of quantum coherence. On the contrary, only a few distinct effective frequencies involves in the time evolution of coherence factor using the ``approximate'' models. Thus, the coherence factor does not relax, and the collapses and revivals of coherence will repeatedly occur.

\begin{figure}[ht]
\centering
\includegraphics[height=11cm]{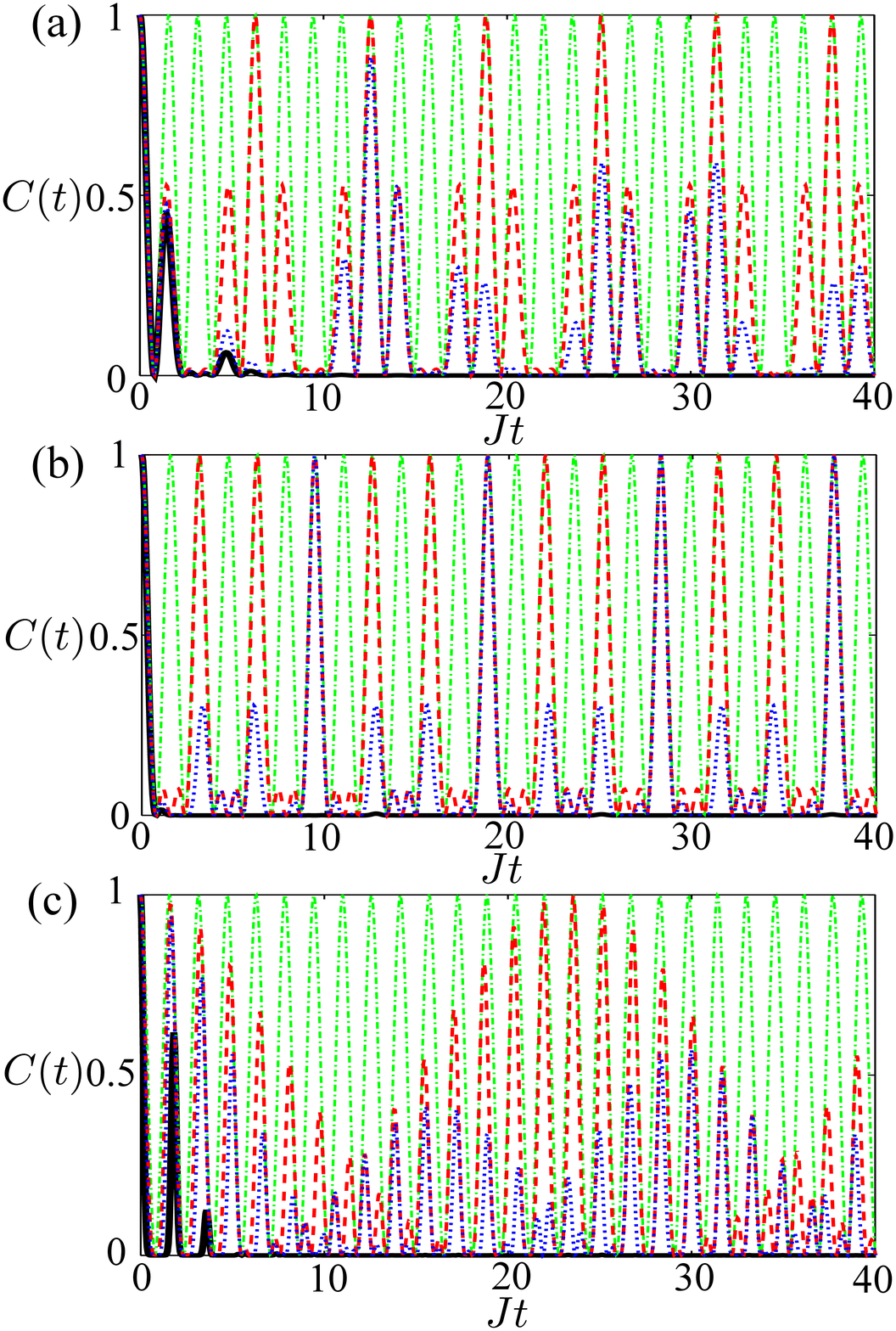}
\caption{ \label{calpha_truncation} (Color online) Coherence factor 
versus time, where the tenth ion is studied and $N=20$.
The different interaction parameters are shown:
(a), $\alpha=2$, (b) $\alpha=1$ and (c) $\alpha=0.1$.
In each figure, the different types of interaction are shown: exact long-range interaction (black solid), next-next-nearest neighbor (blue dotted),
next-nearest neighbor (red dashed) and nearest neighbor (green dot-dashed), respectively.
}
\end{figure}

\begin{figure}[ht]
\centering
\includegraphics[height=14cm]{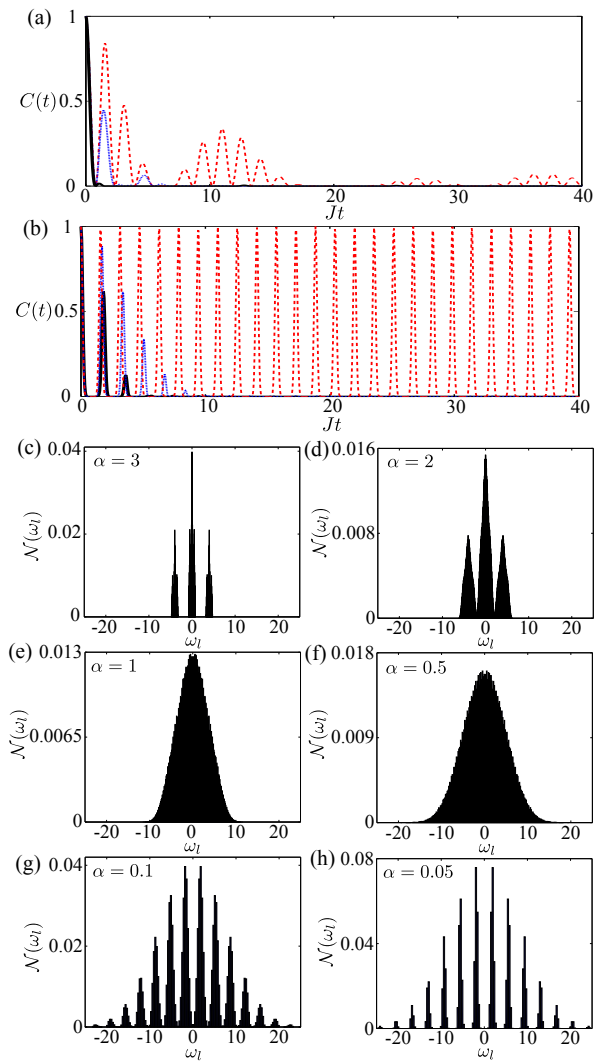}
\caption{ \label{calpha} (Color online) Coherence factor 
versus time in (a) and (b), where the coherence dynamics of the tenth ion is studied and $N=20$. 
The different lines are used to represent the
different parameters $\alpha$: in (a) $\alpha=3$ (red dashed), $\alpha=2$ (blue dotted) and $\alpha=1$ (black solid); in (b) $\alpha=0.1$ (black solid), $\alpha=0.05$ (blue dotted) and $\alpha=0$ (red dashed).
Histograms of the effective frequencies $\omega_k$ are shown in the lower part. The different interaction parameters $\alpha$ are shown: (c) $\alpha=3$, (d) $\alpha=2$, (e) $\alpha=1$, (f) $\alpha=0.5$, (g) $\alpha=0.1$ and (h) $\alpha=0.05$, respectively.
}
\end{figure}

To facilitate our subsequent discussion, we define the relaxation time $t_r$. Within a period, the coherence factor $C(t_r)$ decrease to its initial value being divided by $e$, i.e., $C(0)/e$, where $e$ is equal to the Euler's number. The relaxation rate is equal to $1/t_r$. Here we have assumed that coherence of the system will not resurrect in that period.

In Fig.~\ref{calpha_truncation}, we compare the coherence dynamics of the exact and ``approximate'' models, for the different interaction ranges.
The spin undergoes the relaxation dynamics if the exact model is used. Comparing with the two dynamics in Figs.~\ref{calpha_truncation}(a) and (b), the system, with a smaller $\alpha$, relaxes more rapidly.
When $\alpha=0.1$ is close to zero in Fig.~\ref{calpha_truncation}(c), the spin takes a longer time for relaxation. 
On the contrary, the spin does not relax in the cases of using ``approximation'' models. Similarly, the ``approximate'' model can give out a better approximation of the short-time dynamics of the exact model if the interactions between more distant neighbors are included. 

We further investigate the effect of interaction range on the coherence dynamics. In Fig.~\ref{calpha}, we plot
the coherence factors versus time for the different interaction ranges by using the exact model only. When the range of interaction increases with a smaller $\alpha$ ($\alpha=3,2$ and 1) in Fig.~\ref{calpha}(a), the rate of relaxation increases. However, as $\alpha$ approaches zero, the coherence dynamics behaviors very differently. In Fig.~\ref{calpha}(b), the coherence factor decays more slowly (more revivals before vanishing) if $\alpha$ is smaller. This means that the slower relaxation rate when approaching to the infinite interaction range.
When $\alpha=0$, the coherence factor oscillates periodically. This shows that the behaviors of relaxation dynamics greatly depends
on the interaction range.

In Figs.~\ref{calpha}(c)-(h), we plot the histograms of the effective frequencies $\omega_l$, for the different interaction ranges. Different ranges of interaction exhibit different patterns in the histograms. When $\alpha=3$, the histogram shows three sharp peaks only in Fig.~\ref{calpha}(c). In Fig.~\ref{calpha}(d), the width of peaks become wider if $\alpha=2$. As $\alpha$ further decreases to 1 and 0.5, a single peak is shown in both figures (e) and (f). The peak is wider for the smaller $\alpha$. However, when $\alpha$ is close to 0, several sharp peaks are shown in Figs.~\ref{calpha}(g) and (h). The peaks becomes sharper when the range of interaction is longer. This means that more effective eigenfrequencies are degenerate as the interaction range is close to infinity. Thus, the relaxation does not occur because the system become completely degenerate in the limit of infinite interaction range.

\subsection{Relaxation of block  of spins}
We study the dynamics of quantum coherence of a block of spins with the same initial condition in Eq.~(\ref{zinitial}). Now we consider the
``inside'' system which contains more than one ion. Let us call $N_I$ be the number of spins in the ``inside'' system and the block of spins is located at the centre of a chain. 

\begin{figure}[ht]
\centering
\includegraphics[height=12cm]{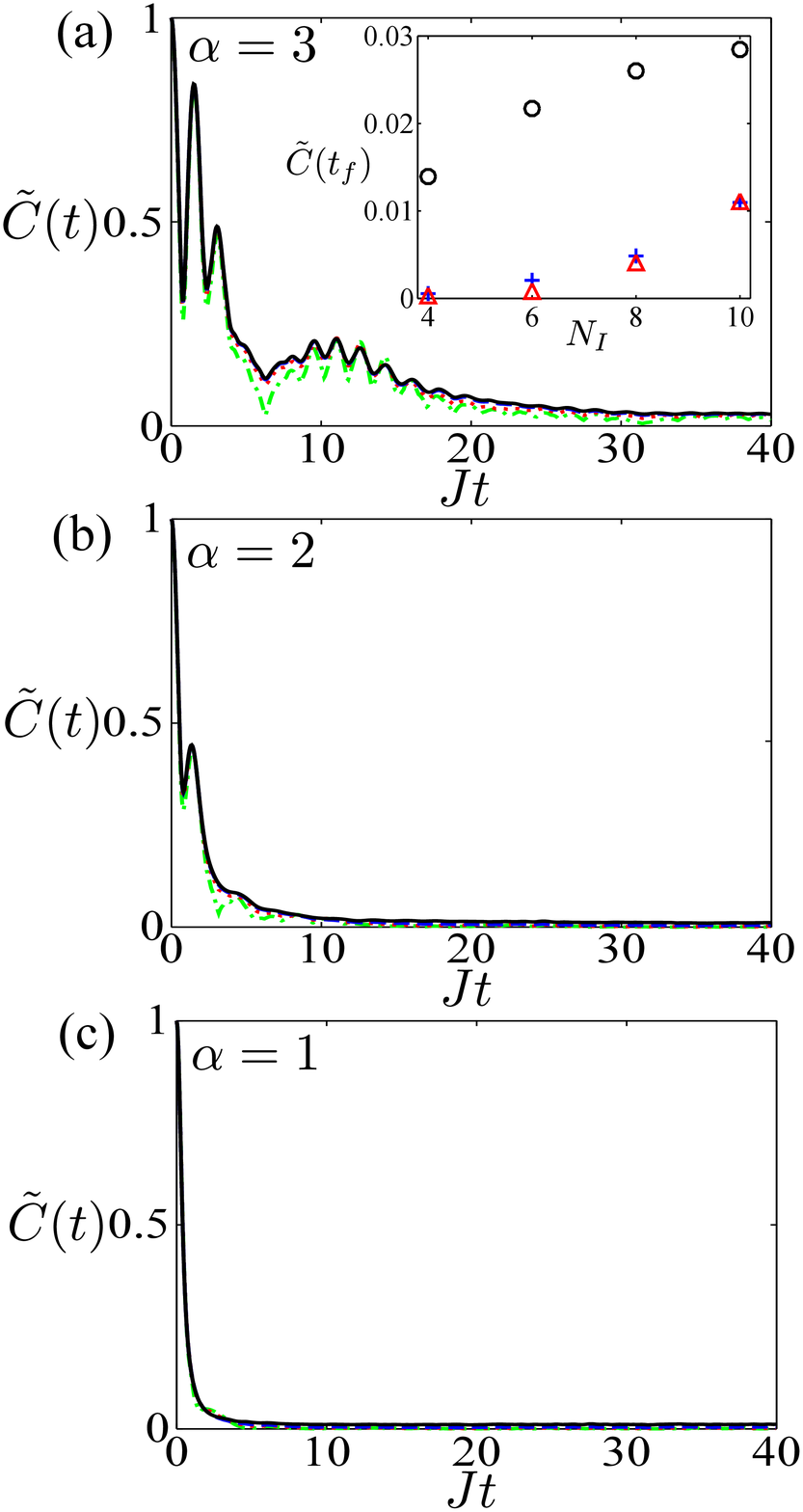}
\caption{ \label{calpha_rdm} (Color online) Normalized coherence factor $\tilde{C}(t)$ versus time, for the different parameters $\alpha$ and $N=20$. The different
parameters are used: (a) $\alpha=3$, (b) $\alpha=2$ and (c) $\alpha=1$. The different lines are used to denote the different sizes of blocks:
$N_I=10$ (black solid), 8 (blue dashed), 6 (red dotted)
and 4 (green dash-dotted) respectively. In the inset, 
the normalized coherence factors $\tilde{C}(t_f)$ are plotted versus $N_I$, where $t_f=40J^{-1}$. The different
parameters $\alpha$ are represented by different symbols:
$\alpha=3$ (black circle), 2 (blue plus) and 1 (red upper triangle), respectively.
}
\end{figure} 

For a larger size of block, it has a higher degree of initial quantum coherence $C(0)$. In fact, $C(0)$
can be evaluated by summing up all the off-diagonal elements, where the density matrix elements of the initial state are all equal.
The initial quantum coherence $C(0)$ is thus given by
\begin{eqnarray}
C(0)&=&\frac{2}{2^{N_I}}\sum^{2^{N_I}-1}_{k=1}k,\\
&=&2^{N_I}-1.
\end{eqnarray}
We define the normalized coherence factor $\tilde{C}(t)$ as 
\begin{eqnarray}
\tilde{C}(t)=\frac{C(t)}{C(0)}.
\end{eqnarray} 
It is convenient to compare the relaxation rates for different block sizes.

In Fig.~\ref{calpha_rdm}, we plot the normalized coherence factors versus time for different sizes of block and range parameters. 
We compare the time evolution of $\tilde{C}$ for the
different $N_I$ with the same $\alpha$ in each subfigure.
The relaxation rate is similar for the different sizes of
blocks. They all decrease and become steady in a long time. The system has a higher relaxation rate when $\alpha$ is smaller. Also, the larger size of block has a higher degree of coherence that remain in the steady state as shown in the inset of Fig.~\ref{calpha_rdm}.

\section{Discussion}
Let us briefly discuss the observation of relaxation dynamics with a long-range Ising model in an experiment.
From the recent experiments of long-range Ising model using trapped ions, the typical value of the spin-spin coupling $J$ \cite{Richerme} is about 100-500 Hz. The coherence lifetime of system \cite{Richerme,Jurcevic} is about 5-10 ms. Given these parameters, the period $\tau$ of coherent dynamics is about 0.5-5$J^{-1}$. This period $\tau$ is sufficiently long to observe the relaxation dynamics due to the long-range interaction. 

Indeed, the long-range spin-spin couplings are generated by off-resonantly coupling to the collective motions of the string via a laser beam. When the transverse magnetic field is absent, the spins can be effectively decoupled from the phonon modes \cite{Wang}. However, the phonon excitations will arise \cite{Wang} in the presence of the
transverse magnetic field. The large magnetic field is required to turn on shortly to prepare the initial state and it is suddenly quenched to initiate the non-equilibrium dynamics. The system then evolves in the absence of the transverse magnetic field. To avoid the intrinsic phonon effects, the laser should be largely detuned from the phonon modes \cite{Wang}. Since the transverse field is turned on initially, the phonon population can be kept to be low during the entire time-evolution.

In addition, the coherence factor of a single spin can be determined from the density matrix of a single spin which can be obtained by just measuring the spin components.
Two-spin density matrix can be obtained by performing quantum state tomography \cite{Christandl,Sugiyama}. In fact, it has been demonstrated in a recent experiment \cite{Jurcevic}. However, it may be more challenging to construct the density matrix as the size of block spin goes large. The dimension of state space will exponentially increase with an increasing number of spins, and thus the complete characterization of multi-spin density matrix becomes hard. Therefore, the study of coherence may limit to a few spins only.

\section{Conclusion}
In summary, we have studied the dynamics of quantum coherence of a single spin and block of spins in a chain of trapped ions, where the ions interact with each other via long-range interactions. The degree of quantum coherence of a single spin rapidly decreases and nearly vanishes in the steady state. Relaxation exhibits in this long-range Ising chain. We have compared the dynamics of the exact long-range Ising model with the ``approximate'' models, where the interactions between the distant ions are truncated in the ``approximate'' models.
Our numerical result suggests that the ``approximate'' model cannot show the relaxation dynamics. In addition, we have examined the effect of long-range interaction on the coherence dynamics. The relaxation rate becomes higher for a longer interaction range. But the spin takes a longer time to relax if the interaction range is in the vicinity of infinite range. We have also investigated the dynamics of coherence of a block of spins. Our study provides a deeper insight of the relationship between the long-range interaction and the quantum-coherence dynamics of a many-body system.

\begin{acknowledgments}
This work was supported in part by the 
National Basic Research Program of 
China Grant No. 2011CBA00300 and No. 2011CBA00301, 
and the National Natural Science Foundation of 
China Grant No.~11304178, No.~61061130540, and No.~61361136003.
\end{acknowledgments}

\appendix


\begin{thebibliography}{99}
\bibitem{Friedenauer}
A. Friedenauer, H. Schmitz, J. T. Glueckert, D. Porras
and T. Schaetz, Nat. Phys. {\bf 4}, 758 (2008).


\bibitem{Kim}
K. Kim {\it et al.}, Nature {\bf 465}, 590 (2010).

\bibitem{Islam}
R. Islam {\it et al.}, Science {\bf 340}, 583 (2013).

\bibitem{Britton}
J. W. Britton {\it et al.}, Nature {\bf 484}, 489 (2012).

\bibitem{Laflorencie}
N. Laflorencie, I. Affleck, M. Berciu,  J. Stat. Mech., P12001 (2005).

\bibitem{Sandvik}
A. W. Sandvik, Phys. Rev. Lett. \textbf{104}, 137204 (2010).

\bibitem{Hauke}
P. Hauke and L. Tagliacozzo, Phys. Rev. Lett. {\bf 111}, 207202 (2013).


\bibitem{Richerme}
P. Richerme {\it et al.}, Nature {\bf 511}, 198 (2014).

\bibitem{Jurcevic}
P. Jurcevic {\it et al.}, Nature {\bf 511}, 202 (2014).

\bibitem{Lieb}
E. Lieb and D. Robinson, Commun. Math. Phys. {\bf 28}, 251 (1972).


\bibitem{Polkovnikov}
A. Polkovnikov, K. Sengupta, A. Silva and M. Vengalattore, Rev. Mod. Phys. {\bf 83}, 863 (2011).


\bibitem{Eisert}
J. Eisert, M. Friesdorf and C. Gogolin, Nat. Phys. {\bf 11}, 124 (2015).


\bibitem{Langen}
T. Langen, R. Geiger, J. Schmiedmayer, Annu. Rev. Condens. Matter Phys. \textbf{6} 201 (2015).

\bibitem{Gogolin}
C. Gogolin and J. Eisert, arXiv:1503.07538.

\bibitem{Deutsch}
J. M. Deutsch, Phys. Rev. A {\bf 43}, 2046 (1991).

\bibitem{Srednicki}
M. Srednicki, Phys. Rev. E {\bf 50}, 888 (1994).

\bibitem{Rigol}
M. Rigol, V. Dunjko and M. Olshanii, Nature {\bf 452}, 854 (2008).

\bibitem{Goold}
J. Goold, M. Huber, A. Riera, L. del Rio, P. Skrzypczyk,
arXiv:1505.07835.

\bibitem{Banuls}
M. C. Ba\~{n}uls, J. I. Cirac, M. B. Hastings, Phys. Rev.
Lett. \textbf{106}, 050405 (2011).

\bibitem{Ng}
H. T. Ng, Phys. Rev. A {\bf 92}, 043634 (2015).


\bibitem{Baumgratz}
T. Baumgratz, M. Cramer, M. B. Plenio, Phys. Rev. Lett. {\bf 113}, 140401 (2014).

\bibitem{Nielsen}
M. A. Nielsen and I. L. Chuang, {\it Quantum Computation and Quantum Information} (Cambridge University Press, Cambridge, 2001).

\bibitem{Romero}
E. Romero {\it et al.}, Nat. Phys. {\bf 10}, 676 (2014).

\bibitem{Koffel}
T. Koffel, M. Lewenstein and L. Tagliacozzo, Phys. Rev. Lett. \textbf{109}, 267203 (2012).

\bibitem{Feig}
M. Foss-Feig, K. R. A. Hazzard, J. J. Bollinger and A. M. Rey, Phys. Rev. A \textbf{87}, 042101 (2013).


\bibitem{Schachenmayer}
J. Schachenmayer, B. P. Lanyon, C. F. Roos and A. J. Daley, Phys. Rev. X \textbf{3}, 031015 (2013). 

\bibitem{Gong}
Z.-X. Gong and L.-M. Duan, New J. Phys. \textbf{15} 113051 (2013).


\bibitem{Sachdev}
S. Sachdev, \textit{Quantum Phase Transitions} (Cambridge University Press, Cambridge, 2001).

\bibitem{Porras}
D. Porras and J. I. Cirac, Phys. Rev. Lett. {\bf 92},
207901 (2004).

\bibitem{Wang}
C.-C. J. Wang  and J. K. Freericks, Phys. Rev. A {\bf 86}, 032329 (2012).

\bibitem{Christandl}
M. Christandl and R. Renner, Phys. Rev. Lett. {\bf 109}, 120403 (2012).

\bibitem{Sugiyama}
T. Sugiyama, P. S. Turner and M. Murao, Phys. Rev. Lett. {\bf 111}, 160406 (2013).





\end{thebibliography}
\end{document}